\documentclass[%
 aip,
 jmp,
 amsmath,amssymb,
reprint, onecolumn 
]{revtex4-1}
\usepackage[english]{babel}
\usepackage{graphicx}
\usepackage{dcolumn}
\usepackage{bm}
\usepackage[utf8]{inputenc}
\usepackage[T1]{fontenc}
\usepackage{mathptmx}
\usepackage{etoolbox}

\makeatletter
\def\@email#1#2{%
 \endgroup
 \patchcmd{\titleblock@produce}
  {\frontmatter@RRAPformat}
  {\frontmatter@RRAPformat{\produce@RRAP{*#1\href{mailto:#2}{#2}}}\frontmatter@RRAPformat}
  {}{}
}%
\makeatother
\begin{document}

\preprint{AIP/123-QED}

\title{Trivalent Feynman Diagrams as a Flag}
\author{Lili Yang}
\email{yanglli6@mail2.sysu.edu.cn}
\affiliation{School of Physics and Astronomy, Sun Yat-Sen University, Zhuhai 519082, China}

\date{\today}

\begin{abstract}
By identifying each standard flag with a trivalent Feynman diagram, the corresponding propagators can be read directly from the flag itself. Within the flag representation, the kinematic Jacobi identity—equivalently, the residue theorem on moduli spaces—admits a natural interpretation as the equivalence between a complete flag and its gapped counterpart. Using flags together with Orlik–Solomon algebras, we reconstruct the intersection numbers of twisted cocycles, thereby obtaining the bi-adjoint amplitude.  Moreover, employing flag simplices enables the construction of the $Z$-amplitude in the $\alpha^{\prime} \to 0$. By further examining pairings of specific flags, we also recover the Cachazo–He–Yuan (CHY) representation of the bi-adjoint amplitude.
\end{abstract}

\maketitle

\section{introduction }
The concept of color–kinematics duality was first discovered by Bern, Carrasco, and Johansson (BCJ) \cite{Bern_2008,Bern_2010}, offering profound insights into the relationship between gravity and gauge theory amplitudes. This duality implies that a gravity amplitude can be expressed as the square of a Yang–Mills amplitude, a correspondence commonly referred to as the double copy. Remarkably, this double-copy structure also emerges in classical gravitational theory \cite{Monteiro_2014,Luna_2015,Ridgway_2016}.This compact formulation has motivated further exploration of the underlying mathematical structures hidden within scattering amplitudes. Over the years, studies of scattering amplitudes have unveiled increasingly rich mathematical patterns that are not apparent at the level of individual Feynman diagrams. These structures exhibit deep connections to both string theory and field theory amplitudes.At tree level, it has been proven that a gravitational amplitude can be obtained by squaring the Yang–Mills amplitude and multiplying by the momentum kernel \cite{Bjerrum_Bohr_2010,Bjerrum_Bohr_2010yang}. In string theory, the BCJ relations can be understood as arising from the $\alpha^\prime \rightarrow 0$ limit \cite{Bjerrum_Bohr_2009} of the monodromy relations in string theory amplitudes.With the discovery of the Cachazo-He-Yuan (CHY) scattering equations \cite{Cachazo_2014,Cachazo_2014563,Cachazo_2014782}, the double copy structure becomes more compact and can be generalized to a broader range of theories. For instance, amplitudes in Einstein-Yang-Mills theory can be expressed in a more concise form using scattering equations \cite{Cachazo_2015,Nandan_2016}. Cachazo, He, and Yuan \cite{Cachazo_2014563} also pointed out that the inverse of the bi-adjoint scalar amplitude (i.e., with two sets of color factors) serves as the momentum kernel of the Kawai-Lewellen-Tye (KLT) relations \cite{Kawai:1985xq} (see also \cite{Mizera_2017,Bjerrum_Bohr_2011}). Scattering amplitudes continue to exhibit increasingly diverse mathematical properties.

Based on the analytic continuation of amplitudes, Britto, Cachazo, Feng, and Witten (BCFW) \cite{PhysRevLett.94.181602} proposed an on-shell recursion relation and used it to directly prove that gravitational amplitudes are the square of Yang-Mills amplitudes. Arkani-Hamed, Bai, He, and Yan introduced the notion of scattering forms linked to field theory \cite{Arkani_Hamed_2018}, and extending this formalism to string theory amplitudes elucidates the KLT relations \cite{arkanihamed2021stringy}. The amplituhedron, a geometric structure related to $\mathcal{N} = 4$ super Yang-Mills (SYM) amplitudes, was discovered in \cite{arkanihamed2014scattering,H.Elvang}, where tree-level amplitudes correspond to volumes of polytopes in a certain auxiliary space. Arkani-Hamed and collaborators revealed deep connections between scattering amplitudes and positive Grassmannian geometry \cite{Arkani_Hamed_2012,Arkani_Hamed_20142,Arkani_Hamed_2014}. Fu and Wang \cite{Fu_2018,Fu_2020} employed screening vertex operators to introduce an algebraic perspective on the Knizhnik-Zamolodchikov (KZ) equations \cite{KNIZHNIK198483} and Z-theory amplitudes \cite{Broedel_2014}. Building on work by Cho, Yoshida, Kita, and Matsumoto developed intersection theory in terms of twisted homology and cohomology \cite{Mimachi_2003, Yoshida, zbMATH01270294,https://doi.org/10.1002/mana.19941680111,Cho_Matsumoto_1995}. Mizera \cite{Mizera_2017I,Mizera_2017C,Mizera_2018} elucidated the mathematical aspects of relations between string theory and field theory amplitudes, demonstrating that the KLT relations are equivalent to twisted period relations discovered by Cho and Matsumoto \cite{Cho_Matsumoto_1995}. Bjerrum-Bohr, Damgaard, Tourkine, and Vanhove \cite{Bjerrum_Bohr_2009} showed that monodromy relations in string theory amplitudes reduce to the Kleiss-Kuijf relations and BCJ duality in field theory in the limit $\alpha^\prime \rightarrow 0$. The underlying structure of monodromy relations in moduli spaces is captured by twisted homology theory \cite{Casali_2019} (see Aomoto's foundational work \cite{10.1093/qmath/38.4.385}).

\textit{We organize the paper as follows.}  
In Section~\ref{section 2}, we introduce the properties of arrangements of hyperplanes and their associated flags, building on the work of Varchenko and Schechtman \cite{Schechtman,varchenko2016critical}. We also provide a brief review of scattering amplitudes. In Section~\ref{section 3}, we identify flags with trivalent Feynman diagrams in the moduli space $\mathcal{M}_{0,n}$. Using the identity of a flag with a gap, we reformulate the residue theorem, which provides a geometric interpretation of the BCJ relations. In Section~\ref{section 4}, we exploit the duality between hyperplanes and flags, together with the bilinear form defined on flags, to reformulate the intersection numbers of twisted logarithmic $p$-forms and obtain the bi-adjoint scalar amplitude. Furthermore, by assigning each flag $F(L^1 \supset \cdots \supset L^p)$ to a singular $p$-simplex, we recover the Z-theory amplitude in the $\alpha^{\prime} \rightarrow 0$ limit. By mapping each critical point $u$ to a special flag $F(u)$, we show that projecting the bilinear form onto these special flags yields the CHY formula for the bi-adjoint scalar amplitude. We present our conclusions in Section~\ref{section 5}. Appendix~\ref{appendix} contains detailed calculations for the five-point example omitted in Section~\ref{subsection 4}.  

\section{Preliminaries}\label{section 2}
\subsection{Scattering Amplitudes }
\label{sec:intro}
In 1986, Kawai, Lewellen, and Tye \cite{Kawai:1985xq}discovered that closed-string amplitudes can be expressed as quadratic combinations of open-string amplitudes. Written in terms of the bi-adjoint scalar amplitude \cite{Mizera_2017}, the KLT relation takes the form
\begin{equation}
    A^{\text{closed}} 
    = \sum_{\beta,\,\gamma} 
      A^{\text{open}}(\beta) \,
      m^{-1}_{\alpha^{\prime}}(\beta \vert \gamma) \,
      A^{\text{open}}(\gamma) \, ,
\end{equation}
where $m^{-1}_{\alpha^{\prime}}$ denotes the inverse of the bi-adjoint scalar amplitude matrix.
Cachazo, He, and Yuan (CHY) \cite{Cachazo_2014563} later identified the KLT kernel with the inverse matrix of the bi-adjoint scalar amplitude. In the field-theory limit $\alpha' \to 0$, the KLT relation becomes
\begin{equation}
\begin{aligned}
M^{\text{GR}} = \sum_{\beta,\gamma} A^{\text{YM}}(\beta) m^{-1} (\beta \vert \gamma ) A^{\text{YM}}(\gamma).
\end{aligned}
\end{equation}
where $m(\beta\vert\gamma)$ is the bi-adjoint scalar amplitude \cite{Cachazo_2014563,Bjerrum_Bohr_2012}. The bi-adjoint scalar amplitude admits a representation as a sum over trivalent Feynman graphs:
\begin{equation}
    \begin{aligned}
    \label{m}
        m(\beta \vert \gamma):=(-1)^{w(\beta \mid \gamma)+1} \sum_{\mathcal{T} \in \mathcal{G}_{\beta} \cap \mathcal{G}_{\gamma}} \frac{1}{\prod_{e \in \mathcal{T}} s_{e}},
    \end{aligned}
\end{equation}
where $s_{e}$ denotes the propagator. Equivalently, this quantity can be expressed as the intersection number of the Parke–Taylor forms with orderings $\beta$ and $\gamma$\cite{zbMATH01270294}:
\begin{equation}
    m(\beta \vert \gamma) := \langle PT(\beta) , PT(\gamma) \rangle.
\end{equation}
Mafra, Schlotterer, and Stieberger showed that open-string partial amplitudes admit an expansion in a basis of $Z$-amplitudes \cite{MAFRA2013461,MAFRA2013419}:
\begin{equation}
    A^{\mathrm{open}}(\beta) 
    = \sum_{\gamma} n(\gamma) \, Z_{\beta}(\gamma) \, ,
\end{equation}
where the $Z$-integrals are defined on the domain $\mathfrak{D}(\beta)$ by
\begin{equation}
    \begin{aligned}
        Z_{\beta}(\gamma):=\int_{\mathfrak{D}(\beta)} \frac{d^{n} z}{\operatorname{vol~SL}(2, \mathbb{R})} \frac{\prod_{i<j}\left(z_{\beta(j)}-z_{\beta(i)}\right)^{\alpha^{\prime} s_{\beta(i), \beta(j)}}}{\left(z_{\gamma(1)}-z_{\gamma(2)}\right)\left(z_{\gamma(2)}-z_{\gamma(3)}\right) \cdots\left(z_{\gamma(n)}-z_{\gamma(1)}\right)}.
    \end{aligned}
\end{equation}
The integrand is the Koba–Nielsen factor times a Parke–Taylor factor \cite{AOMOTO1997119,PhysRevLett.56.2459}. Mizera interpreted these integrals as pairings of twisted cycles and twisted cocycles \cite{Mizera_2017C}. In the field-theory limit one recovers the KLT kernel:
\begin{equation}
    \lim_{\alpha^{\prime} \rightarrow 0} Z_{\beta}(\gamma) 
    = \lim_{\alpha^{\prime} \rightarrow 0} m_{\alpha^{\prime}}(\beta \vert \gamma) 
    = \alpha^{\prime\, 3-n} \, m(\beta \vert \gamma) \, .
\end{equation}

\subsection{Hyperplanes Arrangements and Flag}
This subsection gives an informal introduction to flags and their basic properties for readers who may not be familiar with these topics \cite{Schechtman,Varchenko_2011}.

Let an arrangement be a collection of hyperplanes in the complex space $\mathbb{C}^k$. Denote the arrangement by $\xi=(H_j)_{j\in J}$ with $J={1,\dots,n}$. The complement of the union of all hyperplanes is
\begin{equation}
    U(\xi)=\mathbb{C}^k-\bigcup_{j\subseteq J}H_j.
\end{equation}
An edge $L^\alpha\subset\mathbb{C}^k$ is a nonempty intersection of some hyperplanes; its complex codimension is $\ell_\alpha=\operatorname{codim}_{\mathbb{C}k} L^\alpha$. We set $L^0=\mathbb{C}^k$. The codimension of a hyperplane in this space is $1$,and a vertex corresponds to an edge of maximal codimension, typically with $\ell=k$.

We say hyperplanes $(H_1,\dots,H_p)$ are in general position if $\operatorname{codim}(H_1\cap\cdots\cap H_p)=p$. For a generic arrangement $\xi$, the abelian group $\mathcal{A}^p$ is generated by $p$-tuples $(H_1,H_2,\ldots,H_p)$ with the following relations:
\begin{enumerate}
\item \label{item:gp1} $(H_1,H_2,\ldots,H_p) = 0$ if they are not in general position; that is, if the intersection $H_1\cap \cdots \cap H_p$ is empty or has codimension less than $p$;
\item \label{item:gp2} For any permutation \(\sigma\in S_p\),
\[
    (H_{\sigma(1)},\ldots,H_{\sigma(p)})
    =(-1)^{\operatorname{sgn}(\sigma)} (H_1,\ldots,H_p),
\]
where the sign is \(+1\) for even permutations and \(-1\) for odd ones.

\item \label{item:gp3} For any \((p+1)\)-tuple \((H_1,\ldots,H_{p+1})\) that is not in general position but has nonempty intersection,
\[
    \sum_{i=1}^{p+1}(-1)^i
    (H_1,\ldots,\hat{H_i},\ldots,H_{p+1})=0,
\]
where \(\hat{H_i}\) denotes omission of the \(i\)-th hyperplane.
\end{enumerate}
The multiplication
\begin{equation}
    \begin{aligned}
        (H_{j_1}, \cdots, H_{j_p})\cdot (H_{j_{p+1}}, \cdots, H_{j_{p+q}} )  =  (H_{j_1}, \cdots, H_{j_p}, H_{j_{p+1}}, \cdots, H_{j_{p+q}} ),
    \end{aligned}
\end{equation}
endows the direct sum $A^*(\xi)=\bigoplus_p\mathcal{A}^p$ with the structure of the Orlik–Solomon algebra \cite{Orlik1980CombinatoricsAT}. There exists an isomorphism (often denoted by$\iota$) between the combinatorial data of hyperplanes and differential forms, given by
\begin{equation}
    (H_1,\cdots ,H_p)\cong  \iota(H_1,\cdots,H_p)= d log f_{H_1} \wedge  \dots \cdot \wedge dlog f_{H_p},
\end{equation}
where $f_{H_i}=0$ denotes the defining equation of the hyperplane $H_i$.

Given a $p$-tuple of hyperplanes in general position $(H_1,\dots,H_p)$, define
\begin{equation}
    L^{\alpha_1}=H_1, \quad L^{\alpha_2}=H_1 \cap H_2 , \quad \ldots,\quad   L^{\alpha_p}=H_1\cap H_2\cap \cdots \cap H_p.
\end{equation}
where each $L^{\alpha_i}$ satisfies $\operatorname{codim} L^{\alpha_i}=i$.

Let $\mathcal{F}^p$ denote the set of flags F of length $p$ ending at $L^{\alpha_p}$,
\begin{equation}
    F(L^{\alpha_0}\supset L^{\alpha_1} \supset \cdots \supset L^{\alpha_p}),
\end{equation}
The corresponding standard flag is denoted $F(H_1,\dots,H_p)=F_{\alpha_0,\dots,\alpha_p}$; these standard flags form a basis of $\mathcal{F}^p$. The duality pairing between $\mathcal{A}^p$ and $\mathcal{F}^p$ is
\begin{equation}\label{DR}
\left\langle (H_1, \dots, H_p),\, F_{\alpha_0, \dots, \alpha_p} \right\rangle
=
\begin{cases}
\operatorname{sgn}(\sigma), & \text{if } F_{\alpha_0, \dots, \alpha_p} = F(H_{\sigma(1)}, \dots, H_{\sigma(p)}), \\[4pt]
0, & \text{otherwise.}
\end{cases}
\end{equation}
A flag with a gap at level $i$ (for $0<i<p$) is a sequence
\begin{equation}
    \hat{F}(L^{\alpha_0}\supset L^{\alpha_1} \supset \cdots \supset L^{\alpha_{i-1}} \supset L^{\alpha_{i+1}} \supset \cdots \supset L^{\alpha_p}).
\end{equation}
Flags that contain a gap satisfy the "gap relation":
\begin{equation}
    \begin{aligned}
    \label{gap}
        \sum_{L^i, L^{i-1} \supset L^i \supset L^{i+1}} F(L^{\alpha_0}\supset L^{\alpha_1} \supset \cdots \supset L^{\alpha_{i-1}} \supset L^i\supset L^{\alpha_{i+1}} \supset \cdots \supset L^{\alpha})=0.
    \end{aligned}
\end{equation}
There is a boundary operator $d$ on flags given by
\begin{equation}
    \begin{aligned}
        d F(L^{\alpha_0}\supset L^{\alpha_1} \supset \cdots \supset L^{\alpha_p}) = \sum_{L^{\alpha_{p+1}} \subset L^\alpha}F(L^{\alpha_0}\supset L^{\alpha_1} \supset \cdots \supset L^{\alpha_p}\supset L^{\alpha_{p+1}}).
    \end{aligned}
\end{equation}
and one checks $d^2=0$.

Introduce a weight function $a:\xi\to\mathbb{C}$ defined on the set of hyperplanes, and extend it naturally to the set of edges by
\begin{equation}
    \begin{aligned}
        a(L^{\alpha_p}) = \sum_{H\supset L^{\alpha}} a(H).
    \end{aligned}
\end{equation}
For a flag $F(L^{\alpha_0}\supset\cdots\supset L^{\alpha_p})$, its total weight is defined as the product of the weights associated with its constituent edges:
\begin{equation}
    \begin{aligned}
        a(F)=\prod \limits_{i=1}^p a(L^{\alpha_i}).
    \end{aligned}
\end{equation}
In applications to scalar amplitudes one typically identifies $a(H_{ij})$ with the Mandelstam invariant $s_{ij}=k_i\cdot k_j$.

Let 
\begin{equation}
    \omega(a) = \sum_{H \in \xi } a(H) d\log f_H,
\end{equation}
We define the covariant differential $\nabla_{\omega}:\mathcal{A}^p \to \mathcal{A}^{p+1}$ by 
\begin{equation}
    \begin{aligned}\label{covariant differential }
        \nabla_{\omega} (x) = d x + \omega(a) \wedge x.
    \end{aligned}
\end{equation}
Next, define the contravariant map
\begin{equation}
    \begin{aligned}
        S^{(a)} :\mathcal{F}^p \longmapsto \mathcal{A}^p,\ \ F_{\alpha_0,\ldots ,\alpha_p}\ \longmapsto \sum a(H_1) \dots a(H_p) (H_1, \dots H_p),
    \end{aligned}
\end{equation}
where the sum runs over all $p$-tuples $(H_1,\dots,H_p)$ such that $H_i\supset L^{\alpha_i}$ for each $i=1,\dots,p$. Identifying 
$\mathcal{A}^p\simeq(\mathcal{F}^p)^*$ via $S^{(a)}$,  one obtains a bilinear form $S^{(a)}: \mathcal{F}^p \bigotimes \mathcal{F}^p  \longmapsto \mathbb{C}.$ For standard basis flags  $F_1,F_2 \in F$, the bilinear form is defined as
\begin{equation}
    \begin{aligned}\label{S^a}
    S^{a}(F_1,F_2)= \sum_{(i_1,\dots,i_p)\subset I} a_1\cdots a_p \langle (H_1, \dots ,H_p),F_1 \rangle \langle (H_1, \dots ,H_p),F_2 \rangle,
    \end{aligned}
\end{equation}
where the sum runs over all ordered $p$-tuples $(H_1\ldots H_p)$ of hyperplanes in general position. It satisfies the orthogonality relation
\begin{equation}
S^{(a)}\big(F(H_1,\dots,H_p),F(H^{\prime}_1,\dots,H^{\prime}_p)\big)
=\delta{(H_1,\dots,H_p),(H^{\prime}_1,\dots,H^{\prime}_p)}\;a(H_1)\cdots a(H_p),
\end{equation}
so that distinct standard basis flags are orthogonal under $S^{(a)}$, while a standard flag paired with itself yields the product of the corresponding weights.

The contravariant map defines a homomorphism of complexes $S^{(a)}:(F^*(\xi),d)\rightarrow (A^*(\xi),\nabla_{\omega})$. For an edge $L^i$, put
\begin{equation}
\begin{aligned}
S^{(a)}(L^i) = \sum_{H\supset L^i} a(H)\, d\log f_H \in \mathcal{A}^1,
\end{aligned}
\end{equation}
where the sum is over all $H\in \xi$ containing $L^i$. The map on flags is then defined as
\begin{equation}
\begin{aligned}
S^{(a)} (F(L^1\supset \cdots \supset L^p))=S^{(a)} (L^1) \wedge \cdots \wedge S^{(a)}(L^p).
\end{aligned}
\end{equation}
The duality relation \eqref{DR} can also be written as an integral of differential forms \cite{Schechtman}:
\begin{equation}
\begin{aligned}\label{integral cycle}
\langle (H_1,\ldots ,H_p), F\rangle =\int_{c(F)} \iota(H_1,\ldots,H_p)=(-1)^{\operatorname{sgn}(\sigma)},
\end{aligned}
\end{equation}
where $c(F): (S^1)^p \to U$ is a cycle associated to each flag, and $U$ is the complement of the arrangement.

\section{Residue Theorem as Flag with a Gap}\label{section 3}
Mizera \cite{Mizera_2020} reduced the kinematic Jacobi identities to the residue theorem and advocated a geometric viewpoint on color–kinematics duality. In this chapter we study that perspective using flags. Concretely, we regard the ambient complex space as the moduli space of n-punctured genus-zero Riemann spheres $\mathcal{M}_{0,n}$, as defined by Mizera \cite{Mizera_2020A}. Using the $SL(2,\mathbb{C})$ action we fix three punctures, e.g. $(z_1,z_{n-1},z_n)=(0,1,\infty)$, so that a convenient coordinate chart for $\mathcal{M}_{0,n}$ is
\begin{equation}
\begin{aligned}
\mathcal{M}_{0,n}= \{ ( z_2,z_3,\dots,z_{n-1})\in (\mathbb{CP}^1)^{n-3} \vert \, z_i \neq z_j \text{ for all } i\neq j \}.
\end{aligned}
\end{equation}
When two or more marked points on the sphere collide, the limiting stable curve lies on the boundary $\partial\overline{\mathcal{M}}_{0,n}$ of the Deligne–Mumford compactification\cite{DM}; a collision of exactly two points defines a codimension‑one boundary divisor. For heuristic purposes, we denote the collision locus $z_i=z_j$ by the hyperplane $H_{ij}$. In the moduli space, such collisions correspond to a codimension‑one boundary of $\partial\mathcal{M}_{0,n}$. If an additional marked point is brought into collision, the codimension of the intersection increases by one. Repeating this process for up to $(n-3)$ independent collisions yields a maximal‑codimension boundary —namely, the vertices of $\partial\mathcal{M}_{0,n}$—which correspond to trivalent trees with a total of $(n-3)$ propagators. Such sequences of successive collisions are naturally represented by flags in the incidence structure of $\mathcal{M}_{0,n}$. If only $(n-4)$ such collisions are performed, leaving one point unfixed and three marked points $(z_s,z_t,z_u)$ singled out, then there are three distinct local collision configurations corresponding to the $s$-, $t$- and $u$-channels of Feynman diagrams. Under this configuration, the corresponding flags are precisely those exhibiting gaps, and these satisfy relation \eqref{gap}. In the following, we establish the link between this property and the residue theorem.

Mizera \cite{Mizera_2020A} shows that the intersection number of two differential forms can be expressed as a sum of residues on the moduli space:
\begin{equation}
    \begin{aligned}
        \langle \varphi_- \vert \varphi_+ \rangle=\sum_\upsilon \frac{Res_\upsilon (\varphi_-)Res_\upsilon(\varphi_+)}{\prod_{a=1}^{n-3}P^2_{I_a^\upsilon}}.
    \end{aligned}
\end{equation}
Here, the sum runs over all vertices $\upsilon$ of the boundary $\partial\mathcal{M}_{0,n}$, each of which is in one-to-one correspondence with a trivalent Feynman diagram with propagators $P^2_{I_a^\upsilon}$. The numerators are identified with residues of suitable top-degree holomorphic forms as
\begin{equation}
    \begin{aligned}    
&n_{s}=\operatorname{Res}_{z=z_{s}}(\varphi) , n_{t}=\operatorname{Res}_{z=z_{t}}(\varphi),  n_{u}=\operatorname{Res}_{z=z_{u}}(\varphi), \\
&c_{s}=\operatorname{Res}_{z=z_{s}}(\varphi) ,  c_{t}=\operatorname{Res}_{z=z_{t}}(\varphi) , c_{u}=\operatorname{Res}_{z=z_{u}}(\varphi). 
    \end{aligned}
\end{equation}
The kinematic Jacobi identity of BCJ then follows from the residue theorem \cite{Mizera_2020}:
\begin{equation}
    \begin{aligned}\label{BCJ}
        Res_{z={z_s}}(\varphi)+Res_{z={z_t}}(\varphi)+Res_{z={z_u}}(\varphi)=0.
    \end{aligned}
\end{equation}
for any meromorphic form $\varphi$ whose only poles in the relevant patch are at $z_s$, $z_t,$ and $z_u$.

Next, we analyze the above results from the perspective of flags, with each flag corresponding one-to-one to a trivalent Feynman graph.

Since the gap relation for the 4-point case is trivial, we consider first the \textbf{5-point} case. Fix three punctures (for instance legs 1,4,5) so that the moduli coordinates are $z_2,z_3$ and the collision point $z_2=z_3=0$ corresponds to the intersection
\begin{equation}
    L^2 =H_{12}\cap H_{23}\cap H_{13},
\end{equation}
where we use the notation
\begin{equation}
    H_{12}:z_2=0,\quad  H_{23}:z_2-z_3=0,\quad  H_{13}: z_3=0.
\end{equation}

Denote by $d \log f_{ij}$ the basic one-forms associated with the hyperplanes $H_{ij}$ and let $ a_{ij} = a(H_{ij})$ be their corresponding weights. In the Orlik–Solomon algebra $\mathcal{A}^*(\mathcal{C})$(or via the image of flags under $S^{(a)}$), we consider the following elements:
\begin{equation}
    \begin{aligned} 
        & \mathcal{F}^0 : 1, \\ 
        &\mathcal{F}^1: \omega_{12}=a_{12}d\,log\,f_{H_{12}}, \ \omega_{23}=a_{23}d\,log\,f_{H_{23}},\  \omega_{13}=a_{13}d\,log\,f_{H_{13}}.\\ 
        &\mathcal{F}^2: \omega^\prime_{12}=\omega_{12}\wedge(\omega_{13}+\omega_{23}),\  \omega^\prime_{13}=\omega_{13}\wedge(\omega_{23}+\omega_{12}),\ \omega^\prime_{23}=\omega_{23}\wedge(\omega_{12}+\omega_{13}).
    \end{aligned}
\end{equation}
\begin{figure}
\centering
\includegraphics[width=.5\textwidth]{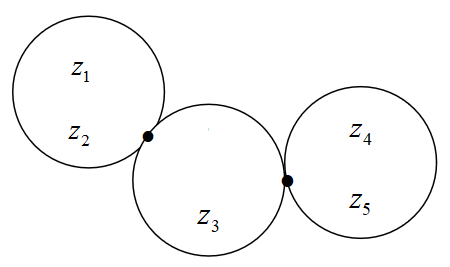}
\qquad
\includegraphics[width=.4\textwidth]{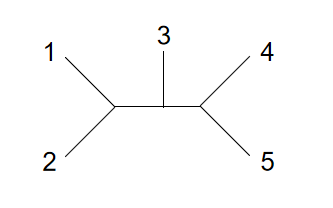}
\includegraphics[width=.5\textwidth]{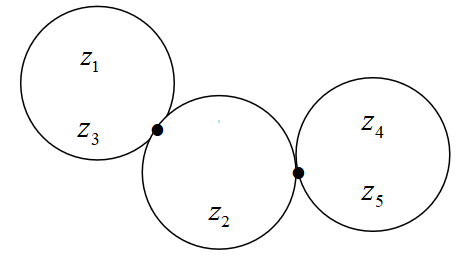}
\qquad
\includegraphics[width=.4\textwidth]{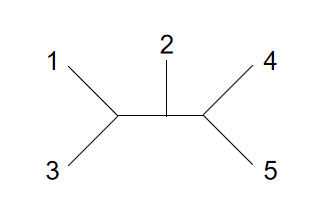}
\includegraphics[width=.5\textwidth]{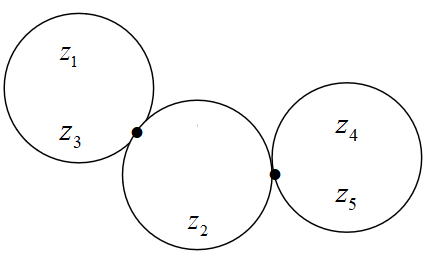}
\qquad
\includegraphics[width=.4\textwidth]{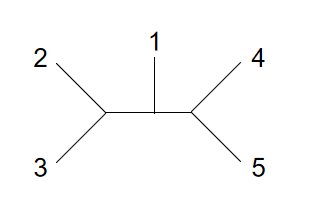}
\caption{The 5-point example illustrating the relation between trivalent tree diagrams and flags generated by the hyperplanes $H_{12},H_{13},H_{23}$.}
\label{fig:5pt}
\end{figure}
The twisted (covariant) differential $\nabla_{\omega} =d +\omega \wedge\cdot $ acts as $\omega_{ij}\mapsto \omega^\prime_{ij}$. These two-forms satisfy the linear relation
\begin{equation}
    \begin{aligned}
\omega^\prime_{12}+\omega^\prime_{13}+\omega^\prime_{23}=0.
    \end{aligned}
\end{equation}
which is the image under $S^{(a)}$ of the flag gap relation (\ref{gap}) for the corresponding flags.Concretely, the three standard flags ending at the point $L^2=(0,0)$ are
\begin{equation}
    \begin{aligned}\label{5pt}
        F_1=F(L^0\supset H_{12}\supset L^2),\ F_2=(L^0\supset H_{13}\supset L^2),\ F_3=(L^0\supset H_{23}\supset L^2).
    \end{aligned}
\end{equation}
Each flag corresponds to a trivalent Feynman graph (see figure \ref{fig:5pt}). 
When we remove a hyperplane corresponding to a codimension-1 edge, 
such as $H_{12}$, $H_{13}$, or $H_{23}$, we obtain a \emph{flag with a gap}. In the 5‑point example, the sum of the complete flags given in 
Eq.~\eqref{gap} corresponds to the choice $i=1$ in the gap equation, i.e.,
 \begin{equation}
    \begin{aligned}
        F_1+F_2+F_3=0.
    \end{aligned}
\end{equation}
By the duality between flags and differential forms, we have
\begin{equation}
    \begin{aligned}
        \langle \varphi,F_1\rangle+\langle \varphi,F_2\rangle+\langle \varphi,F_3\rangle=\langle \varphi,F_1+F_2+F_3 \rangle=0.
    \end{aligned}
\end{equation}
which provides a residue-theorem realization of the Jacobi relation for the three channels in this flag configuration.

\begin{figure}[htbp]
\centering
\includegraphics[width=.5\textwidth]{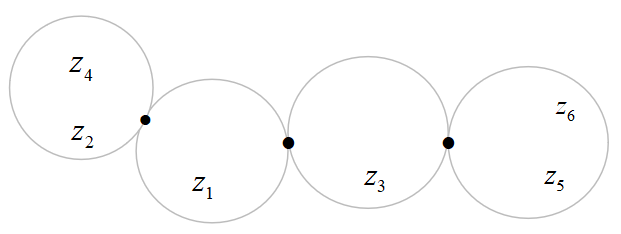}
\qquad
\includegraphics[width=.4\textwidth]{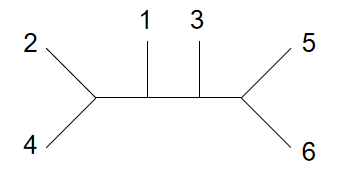}
\includegraphics[width=.5\textwidth]{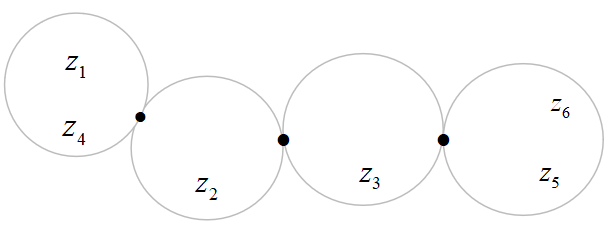}
\qquad
\includegraphics[width=.4\textwidth]{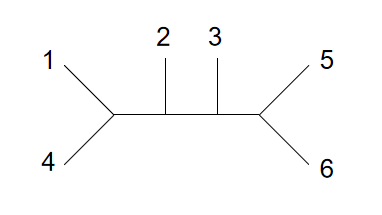}
\includegraphics[width=.5\textwidth]{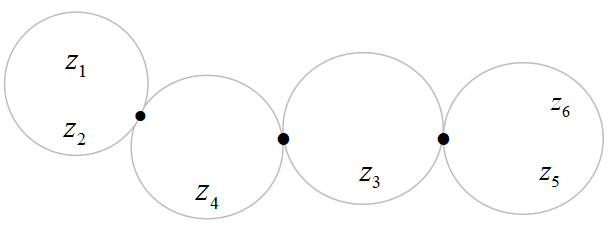}
\qquad
\includegraphics[width=.4\textwidth]{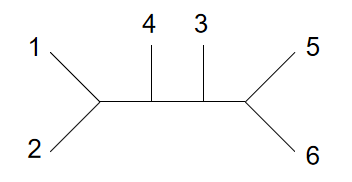}
\caption{The 6-point example illustrating the relation between trivalent tree diagrams and flags generated by the hyperplanes $H_{12},H_{14},H_{24}$.}
\label{fig:6pt}
\end{figure}

\textbf{6-point case.} Fix three punctures (for example, legs 3,5, and 6) so that the remaining coordinates contain $z_1,z_2,z_4$ and consider the local configuration where the point $L^3 =(0,0,0)$ is obtained after further collisions. Denote three hyperplanes that meet along a common axis $L^2$ ($z_3-$axis) by $H_{12},H_{14},H_{24}$, so that any two of these intersect in the same codimension‑2 edge $L^2$ and all three meet at the point $L^3$. There are three flags corresponding to the three planar trivalent Feynman graphs shown in figure~\ref{fig:6pt}. Concretely,
\begin{equation}
    \begin{aligned}
        F_1(L^0\supset H_{14}\supset L^2 \supset L^3),\, F_2(L^0\supset H_{12}\supset L^2 \supset L^3),\, F_3(L^0\supset H_{24}\supset L^2 \supset L^3).
\end{aligned}
\end{equation} 

Their images under the map $S^{(a)}$ in $\mathcal{A}^3(\mathcal{C})$ can be written  as
\begin{equation}
    \begin{aligned}
\omega_1=\omega_{14}\wedge(\omega_{12}+\omega_{24})\wedge(\omega_{13}+\omega_{23}+\omega_{34}),\notag\\
\omega_2=\omega_{12}\wedge (\omega_{24}+\omega_{12})\wedge (\omega_{13}+\omega_{23}+\omega_{34}),\notag\\
\omega_3=\omega_{24}\wedge(\omega_{14}+\omega_{12})\wedge(\omega_{13}+\omega_{23}+\omega_{34}),
    \end{aligned}
\end{equation}
and satisfy 
\begin{equation}
    \begin{aligned}
        \omega_1+\omega_2+\omega_3=0.
    \end{aligned}
\end{equation}
Correspondingly, the equation for a flag with a gap in the case $i=1$ (with $H_{12}$, $H_{14}$, $H_{24}$ omitted; see Eq.~\eqref{gap}) satisfies the following condition:
\begin{equation}
    \begin{aligned}
        F_1+F_2+F_3=0.
    \end{aligned}
\end{equation}
There is also another gap choice (corresponding to the situation where $i=2$  is selected in formula~\eqref{gap}). For instance, by fixing the intermediate hyperplane $L^1 = H_{14}$, the complete refinements of the gap yield the following flags:
\begin{equation}
    \begin{aligned}
        F^{\prime}_1(L^0\supset H_{14}\supset L_1^2 \supset L^3),\,F^{\prime}_2(L^0\supset H_{14}\supset L_2^2 \supset L^3),\,
F^{\prime}_3(L^0\supset H_{14}\supset L_3^2 \supset L^3).
    \end{aligned}
\end{equation}
where $L_1^2$, $L_2^2$ and $L_3^2 $ are the $z_2$-axis, $z_3$-axis, and $H_{14}\cap H_{34}$, respectively. These flags correspond to another set of trivalent tree diagrams, which are equivalent to rotations of the other legs. As the arguments are entirely analogous, we omit the explicit diagrams and formulas. The associated flags satisfy the gap relation
\begin{equation}
    \begin{aligned}
        F^{\prime}_1+F^{\prime}_2+F^{\prime}_3 =0 .
    \end{aligned}
\end{equation} 

\textbf{General case}. If only (n-4) independent collisions are performed \cite{Mizera_2020}—that is, when the configuration is still one residue short of reaching a trivalent factorization—one obtains a flag with a gap,
\begin{equation}
    \hat{F}(L^0\supset L^1\supset \cdots \supset L^{i-1} \supset L^{i+1}\supset\cdots  \supset L^{n-3} ),
\end{equation}
Filling this gap in the three possible ways generates the three channel flags $F_s,F_u,F_t$, which satisfy the gap relation (\ref{gap})
\begin{equation}
    \begin{aligned}
        F_s+F_u+F_t=0.
    \end{aligned}
\end{equation}
Thus, the Jacobi identity (\ref{BCJ}) can be interpreted as a residue theorem applied to the three flags obtained by filling the corresponding gap. In this picture, the Jacobi identity naturally takes the form
\begin{equation}
    \begin{aligned}
        \langle \varphi_{\pm},F_s\rangle+\langle \varphi_{\pm},F_t\rangle+\langle \varphi,F_u\rangle=\langle \varphi_{\pm},F_s+F_t+F_u \rangle=0.
    \end{aligned}
\end{equation}
where the final equality expresses precisely the flag–gap relation.

\section{Flag Representation of the Scalar Amplitude}\label{section 4}
In this section, we present the scalar amplitudes of tree graphs in terms of flag representations. We discuss the corresponding intersection number associated with flags, analyze the $Z-$ amplitudes in the $\alpha^{\prime}\rightarrow 0$ limit using flag simplices, and explore the CHY representations of scalar amplitudes associated with specific flag configurations.

\subsection{Intersection Number of Cocycles Flags and Scalar Amplitudes}

In this section, we express the intersection number of twisted cocycles in terms of flags. Our strategy is to construct local solutions by decomposing logarithmic $p$-forms into the fundamental elements of the Orlik–Solomon algebra $\mathcal{A}^{p}$, and then employing the flags that are dual to these fundamental elements. Finally, we relate the scalar amplitude to its representation in the flag basis.

For logarithmic $p$-forms of the form
\begin{equation}\label{p-form}
\begin{aligned}
&\varphi_I=\frac{d\log f_{H_{i_1}}}{d\log f_{H_{i_2}}}\wedge \frac{d\log f_{H_{i_2}}}{d\log f_{H_{i_3}}}\wedge \cdots \wedge \frac{d\log f_{H_{i_p}}}{d\log f_{H_{i_{p+1}}}},\
& I=(i_1,i_2,\dots,i_{p+1}),
\end{aligned}
\end{equation}
the intersection number of twisted cocycles is defined \cite{zbMATH01270294} as
\begin{equation}
    \begin{aligned}
    \label{IN}
        \langle \varphi_I,\varphi_J^\prime \rangle_\omega =\int \varphi_I\wedge \iota_\omega(\varphi_J^\prime),
    \end{aligned}
\end{equation}
where $\iota_\omega(\varphi_J)$ is a compactly supported $p$-form on $U(\xi)$ vanishing in small neighbourhoods of the singular points $L^p$. This integral can be simplified using Stokes' theorem together with the residue theorem.
A crucial step is to construct a local solution $\psi^{p-1}_m$ on $U(\xi)$ satisfying
\begin{equation}\label{local solution}
    \bigtriangledown_\omega\psi^p_m =\varphi_i,
\end{equation}
where $\varphi_i$ is a linear dlog-form unit corresponding to one of the $p$ hyperplanes chosen from the $p+1$ hyperplanes composing $\varphi_I$. Matsumoto \cite{zbMATH01270294} showed that, near a singular point $L^p$ this local solution is given by the ratio of the residues of $\varphi$ and $\omega$. This result matches precisely the structure of scalar scattering amplitudes\cite{Mizera_2018,Mizera_2020A}.

The logarithmic $p$-form can be expanded in the basis $\iota(H_{i_1},\dots,H_{i_{p+1}})$ of the Orlik–Solomon algebra $\mathcal{A}^{p}$, constructed from hyperplanes $(H_{i_1},\dots,H_{i_{p}})$ in general position. In particular,:
\begin{equation}
    \begin{aligned}
        \varphi_I=\sum_{j=1}^{p+1} (-1)^{p+1} \cdot (-1)^{i} \iota(H_{i_1},H_{i_2},\dots\hat{H}_{i_j},\dots,H_{i_{p+1}}),
    \end{aligned}
\end{equation}
where the hat symbol $\hat{H}_{i_j}$ denotes omission of the corresponding hyperplane, and
\begin{equation}
    \iota(H_{i_1},\dots,H_{i_p}) := d\log f_{H_{i_1}}\wedge\cdots\wedge d\log f_{H_{i_p}}.
\end{equation}
This logarithmic structure naturally identifies the Parke–Taylor form $PT(1,\alpha , n-1, n)$ with ordering $\alpha$\cite{Mizera_2017}. In other words, the PT-form can be expanded in the basis that is dual to the flag basis in $\mathcal{A}^{n-3}$(where $p=n-3$). For instance, consider the hyperplanes $H_{i_1}=H_{12},H_{i_2}=H_{23},H_{i_3}=H_{34}$; any two of these form $2$‑tuples in general position. The corresponding Parke–Taylor form $PT(12345)$ is defined as\cite{Mizera_2017C}
\begin{equation}
    \begin{aligned}
        PT(12345)= (-1) \left(dlog \frac{0-z_2}{z_2-z_3} \wedge dlog \frac{z_2-z_3}{z_3-1}\right).
    \end{aligned}
\end{equation}
This form can be expressed as a linear combination of basis elements in the Orlik–Solomon algebra $\mathcal{A}^2$,
\begin{equation}
    \begin{aligned}\label{PT(12345)}
        PT(12345)= - \iota (H_{12},H_{23})+ \iota (H_{12},H_{34}) -\iota (H_{23},H_{34}),
    \end{aligned}
\end{equation}
For non–general-position $(2{+}1)$‑tuples, such as $(H_{12},H_{23},H_{13})$, the contribution is zero by definition~\ref{item:gp3}, as the three hyperplanes intersect at the same point with codimension~2.

Next, we represent the local solution \eqref{local solution} in terms of a flag structure and derive the corresponding expression for the scalar amplitude associated with that flag.

Introduce the localized 1-forms on a codimension-$i$ edge $L^i$ by
\begin{equation}
    S^{(a)}(L^i) \;=\; \sum_{H \supset L^i} a(H)\, d\log f_H \in \Omega^1,
\end{equation}
so $S^{(a)}(L^p)$ collects the singular part of $\omega$ localized at the max-codimension edge $L^p$.
For a flag $F(L^0\supset L^1\supset\cdots\supset L^p)$ we write the product of these localized 1-forms as
\begin{equation}
    S^{(a)}(F) \;=\; \prod_{i=1}^p S^{(a)}(L^i),
\end{equation}
A local solution associated to a flag $F_m$ has the form
\begin{equation}
    \psi_m^{\,p-1} \;=\; (-1)^{\sigma(m)}\,\frac{S^{(a)}(L^1)\cdots S^{(a)}(L^{p-1})}{S^{(a)}(F_m,F_m)},
\end{equation}
and satisfies
\begin{equation}
    \nabla_\omega \psi_m^{\,p-1}
    \;=\;
    (-1)^{p-1}\, \frac{S^{(a)}(F_m)}{S^{(a)}(F_m,F_m)} \;=\; \varphi_m^p,
\end{equation}
where $\varphi_m^p$ is the full logarithmic $p$-form dual to the flag $F_m$ and $(-1)^{\sigma(m)}$ accounts for permutation/sign conventions of the flag.

The value of $\psi^{p-1}$ at the singular point $L^p$ is given by the ratio between the residues of $\varphi_I$ and $\omega$ at $L^p$:
\begin{equation}\label{solution}
    \psi^{p-1}_m(L^p)\;=\; \frac{\delta(m;I)}{S^{(a)}(F_m,F_m)} ,
\end{equation}
with $\delta(m;I)=\pm 1$ determined by which hyperplanes are chosen in the multi-index $I$ \cite{zbMATH01270294}.

\textbf{Four-point example}. Consider the Parke–Taylor form $PT(1234)$ , which corresponds to the logarithmic one-form $\varphi_I =\iota(H_{12})/\iota(H_{23})$. Since the configuration has two singular points at $z=0$ and $z=1$, the one-form can be decomposed into
\begin{equation}
    \varphi_1 = \iota (H_{12}), \quad \varphi_2 = \iota (H_{23 }),
\end{equation}
where the index set $I=\{1,2\}$ labels the hyperplanes $H_{12}$ and $H_{23}$, respectively.
For each component $\varphi_i$, there exists a local function $\psi_i$, constructed from the corresponding flag, satisfying
\begin{equation}
    \nabla_\omega \psi_i =\varphi_i
\end{equation}
The value of $\psi_i$ on singular points $L^1$ is 
\begin{equation}
    \begin{aligned}
        \psi_i(L^1)= \frac{\delta(i;I)}{a(H_i)},
    \end{aligned}
\end{equation}
where $\delta(i;I)=\delta_{i,1}-\delta_{i,2}$ is the Kronecker symbol determining the sign induced by the choice of multi-index $I$ .

For the flag $F_1 = F(H_{12})$, one has
\begin{equation}
    \begin{aligned}
        \psi_1(L^1) =\frac{1}{a(H_{12})}, \quad \nabla_\omega \psi_1 \rightarrow \frac{S^{(a)} (F_1)}{S^{(a)}(F_1,F_1)}=\frac{a(H_{12})\iota(H_{12})}{a(H_{12})}=\iota(H_{12}).
    \end{aligned}
\end{equation}
Similarly, for $F_2 = F(H_{23})$, one finds
\begin{equation}
    \begin{aligned}
        \psi_2(L^1)= \frac{-1}{a(H_{23})}, \quad \nabla_\omega\psi_2\rightarrow\frac{-S^{(a)} (F_2)}{S^{(a)}(F_2,F_2)}=\frac{a(H_{23})\iota(H_{23})}{a(H_{23})}=-\iota(H_{23}),
    \end{aligned}
\end{equation}
where the overall sign follows from the orientation convention $\delta(\cdot;I)$.

\textbf{Five-point example.}  Locally, the logarithmic two-form corresponding to $PT(12345)$ in Eq.~\eqref{PT(12345)} is
\begin{equation}
    \begin{aligned}
        \varphi_I =\frac{\iota(H_{12})}{\iota(H_{23})}\wedge \frac{\iota(H_{23})}{\iota(H_{34})} 
    \end{aligned}
\end{equation}
As a concrete example, let us consider the boundary singularity at $(0,1)$, for which the corresponding non-vanishing form $ \varphi_I$ is
\begin{equation}
    \varphi_{(0,1)}=\iota(H_{12},H_{34}) (\text{or}\, -\iota(H_{34},H_{12}))
\end{equation}
although the same analysis applies to all other singular points.

In local coordinates $(z_2,z_3)$ near the singular point $(0,1)$. there are two flags associated with this point, $F_1 = F(H_{12}, H_{34})$ and $F_2 = F(H_{34}, H_{12})$, which differ only by an overall sign. 
The two codimension-one edges meeting at this boundary contribute the dual one-forms
\begin{equation}
    \begin{aligned}
        S^a(L_1^1)=a(H_{12})\iota(H_{12}),\ \ S^a(L_2^1)=a(H_{34})\iota(H_{34}),
    \end{aligned}
\end{equation}
The corresponding products along the two flags are
\begin{equation}
    \begin{aligned}
        &S^a(F_1)=S^a(L_1^1)S^a(L_1^2)=(a(H_{12})\iota(H_{12}))(a(H_{12})\iota(H_{12})+a(H_{34})\iota(H_{34})) \\ 
        &S^a(F_2)=S^a(L_2^1)S^a(L_2^2)=(a(H_{34})\iota(H_{34}))(a(H_{34})\iota(H_{34})+a(H_{12})\iota(H_{12})).
    \end{aligned}
\end{equation}
Thus, according to the definition in~\eqref{S^a}, we have $S^{(a)}(F_1,F_1) = S^{(a)}(F_2,F_2) $.
 
There exist two 1-form solutions, $\psi^1_1$ and $\psi^1_2$,
\begin{equation}
    \begin{aligned} 
        &\psi^1_1=\frac{a(H_{12})\iota(H_{12})}{a(H_{12})a(H_{34})}=\frac{\iota(H_{12})}{a(H_{34})},\\ 
        &\psi^1_2=\frac{-a(H_{34}) \iota(H_{34})}{a(H_{12})a(H_{34})}=\frac{-\iota(H_{34})}{a(H_{12})}.
    \end{aligned}
\end{equation}
Both forms satisfy the local differential equation
\begin{equation}
        \nabla_\omega \psi^1_1 = \nabla_\omega \psi^1_2=\varphi_{(0,1)}
\end{equation}
Explicitly, the covariant derivatives evaluate to
\begin{equation}
    \begin{aligned}
        \nabla_\omega \psi^1_1 &=\frac{S^a(L_1^2)S^a(L_1^1)}{S^{(a)}(F_1,F_1)}=\iota(H_{12},H_{34}),
    \end{aligned}
\end{equation}
and similarly
\begin{equation}
    \begin{aligned}
        \nabla_\omega \psi^1_2 =\frac{-S^a(L_2^2)S^a(L_2^1)}{S^{(a)}(F_2,F_2)}=-\iota(H_{34},H_{12}).
    \end{aligned}
\end{equation}
One can show that there exists a unique holomorphic function $\psi_{(0,1)}$ satisfying\cite{zbMATH01270294}
\begin{equation}
    \begin{aligned}
        \nabla_\omega \psi_{(0,1)} = \frac{S^{a}(L^2)}{S^{(a)}(F,F)}=\frac{1}{a(H_{12})a(H_{34})}(a(H_{12})\iota(H_{12})+a(H_{34})\iota(H_{34}))=\psi^1_1 -\psi^1_2, 
    \end{aligned}
\end{equation} 
and whose value at the singular point $L^2$ is 
\begin{equation}
    \psi_{(0,1)}(L^2)=\frac{1}{S^{(a)}(F,F)}=\frac{1}{a(H_{12})a(H_{34})}.
\end{equation}
This result coincides with the expression obtained by Matsumoto~\cite{zbMATH01270294}. More generally, in the $p$dimensional case there exists a unique holomorphic function $\psi_I$ whose values at the singularities satisfy
\begin{equation}
    \begin{aligned}
        \psi_I = \frac{\delta({I,J})}{S^{(a)}(F_I,F_I)},
    \end{aligned}
\end{equation}
where the multi-index $J=(j_0,j_1,j_2,\ldots,j_p)$ corresponds to the $p$-forms $\varphi_J$, and the multi-index $I=(i_0,i_1,i_2,\ldots,i_p)$ specifies a selection of $p$ hyperplanes that form the flag $F^p_m=F(H_{i_0},H_{i_1},\ldots ,H_{i_{p-1}})$. For a complete proof, we refer the reader to Matsumoto’s article\cite{zbMATH01270294}. The main purpose of the present paper is to reformulate these results using flags and to provide a flag-based representation that allows the propagator of a Feynman trivalent graph to be read off directly.

Consequently, in the general case where $\varphi_I\in\mathcal{A}^p$ is a linear combination of basis elements dual to the flags  $F_m$, one has
\begin{equation}
    \langle \varphi_I,F_m\rangle=\delta(m;I),\qquad
    \frac{\langle F_m,\varphi_J\rangle}{S^{(a)}(F_m,F_m)}=\frac{\delta(m;J)}{S^{(a)}(F_m,F_m)}.
\end{equation}
Therefore, the intersection index \eqref{IN} can be represented via the flag expansion as
\begin{equation}\label{R}
    \langle \varphi_I,\varphi_J\rangle_\omega
    \;=\;
    \sum_m \langle \varphi_I,F_m\rangle \; \frac{\langle F_m,\varphi_J\rangle}{S^{(a)}(F_m,F_m)}.
\end{equation}
where the first identity expresses the Kronecker–delta orthogonality between the chosen basis and the dual flag basis.

Replacing $\varphi_I,\varphi_J$ by Parke–Taylor forms gives the bi-adjoint scalar amplitude \cite{Cachazo_2014,Mizera_2018}:
\begin{equation}\label{PT}
    m(\alpha\mid\beta)=\langle PT(\alpha),PT(\beta)\rangle_\omega
    \;=\;
    \sum_m \langle PT(\alpha),F_m\rangle \; \frac{\langle F_m,PT(\beta)\rangle}{S^{(a)}(F_m,F_m)}.
\end{equation}
Each basis flag $F_m$ identifies a trivalent Feynman diagram; the factors $S^{(a)}(F_m,F_m)$ play the role of propagators for that diagram.

\textbf{Four points example.} Consider the ordering $I = (1234)$ and the multi-index $I = (12, 23)$. In this case, equation $(\ref{PT})$ takes the form
\begin{equation}
    \begin{aligned}
        \langle PT(I),PT(I)\rangle&=\frac{\langle PT(I),F_1\rangle \langle PT(I),F_1\rangle }{S^{a}(F_1,F_1)}+\frac{\langle PT(I),F_2\rangle \langle PT(I),F_2\rangle }{S^{a}(F_2,F_2)}\notag \\
&=\frac{1}{s}+\frac{1}{t}.
    \end{aligned}
\end{equation}
where we replace $a(H_{12})$ and $a(H_{23})$ by $s$ and $t$, respectively.
For the ordering $J = (1324)$ with multi‑index $J = (32,\,24)$, 
there is only one flag common to both $I$ and $J$, namely the flag generated by the hyperplane $H_{23}$. Therefore,
\begin{equation}
    \begin{aligned}
        \langle PT(I),PT(1324)\rangle=0+\frac{\langle PT(I),F_2\rangle \langle PT(1324),F_2\rangle }{S^{a}(F_2,F_2)}=-\frac{1}{t}.
    \end{aligned}
\end{equation}
The minus sign originates from the mismatch in the sign factors assigned to the common flag $F_2 = F(H_{23})$, with $\delta(23, J) = 1$ in the case of $J$ and $\delta(23, I) = -1$ in the case of $I$.

\textbf{Five points example.} We consider the PT‑forms $PT(12345)$ and $PT(13245)$. One finds that
\begin{equation}
    \begin{aligned}
    \label{PT2}
        &PT(12345) =-\iota(H_{23},H_{34})+\iota(H_{12},H_{34})-\iota (H_{12},H_{23}),\\
        &PT(13245)=\iota (H_{32},H_{24})-\iota (H_{13},H_{24})+\iota (H_{13},H_{32}).
    \end{aligned}
\end{equation}
There are two degenerate singular points, $(0,0) $ and $(1,1)$. At $ (0,0) $ the hyperplanes $(H_{12},H_{13},H_{23})$ fail to be in general position.  One resolves these triple intersections by a minimal blow‑up (see \cite{Mimachi_20033,De}), introducing the exceptional divisor $H_{123}$ with weight $a(H_{123})=s_{123}=s_{12}+s_{23}+s_{13}$.In the blown‑up space one can choose the standard flags
\begin{equation}
    F_1=F(H_{12}\supset L_{12}^2),\quad F_2=F(H_{23}\supset L_{23}^2),\quad F_3=F(H_{13}\supset L_{13}^2),
\end{equation}
where \(L_{ij}^2=H_{ij}\cap H_{123}\).
 
The two-form $\iota(H_{12},H_{23})$ decomposes on the basis dual to these standard flags as
\begin{equation}
    \begin{aligned}
    \label{23}
        \iota(H_{12},H_{23})= -(\iota(H_{12},H_{123})+\iota(H_{23},H_{123})),
    \end{aligned}
\end{equation}
and similarly
\begin{equation}
    \begin{aligned}
    \label{32}
        \iota(H_{13},H_{32})= -(\iota(H_{13},H_{123})+\iota(H_{23},H_{123})),
    \end{aligned}
\end{equation}
Therefore the only nonzero common flag contributing to $\langle \iota(H_{12},H_{23}),\iota(H_{13},H_{32})\rangle$ is \(F_2=F(H_{23}\supset H_{23}\cap H_{123})\), and one obtains
\begin{equation}
\langle \iota(H_{12},H_{23}),\iota(H_{13},H_{32})\rangle
=\frac{\langle \iota(H_{23},H_{123}),F_2\rangle\langle F_2,\iota(H_{23},H_{123})\rangle}{S^{(a)}(F_2,F_2)}
=-\frac{1}{s_{23}s_{123}}.
\end{equation}

The same analysis at the point $(1,1)$ introduces the hyperplane $H_{234}$, with weight  $s_{234}=s_{24}+s_{23}+s_{34}$ and with decompositions analogous to \eqref{23} and \eqref{32}. 
The only nonzero common flag in this case is \(F_4=F(H_{23}\supset H_{23}\cap H_{234})\), which contributes
\begin{equation}
    \begin{aligned}
        \langle \iota(H_{23},H_{34}),\iota(H_{32},H_{24})\rangle =\frac{-1}{s_{23}s_{234}}.
    \end{aligned}
\end{equation}
Collecting the contributions from both degenerate points yields the intersection number (also denoted \(m(\alpha\mid\beta)\) in \cite{Cachazo_2014563})
\begin{equation}
    \begin{aligned}
       \langle PT(12345),PT(13245) \rangle = - \frac{1}{s_{23}s_{234}} - \frac{1}{s_{34}s_{234}}.
    \end{aligned}
\end{equation}

\subsection{Flag Simplex and Z-Amplitude in Filed Theory Limit}\label{subsection 4}
We choose a point $x(L)\in L^0 $ for every edge $L$. To each flag $F(L^1\supset \cdots \supset L^p)$ we assign a singular $p$-simplex \cite{Schechtman} $\Delta(F): \Delta^p \rightarrow  \mathbb{C}^k$, where the standard oriented $p$-simplex is defined as
\begin{equation}
    \Delta^p =\lbrace(x_0,x_1,x_2,\ldots,x_p)\in \mathbb{R}^{p+1}\vert \ x_i\geq 1; \sum x_i =1\rbrace.
\end{equation}
The vertices of $\Delta^p$ are the standard basis vectors $e_i$ (with a 1 in the $i$-th position and 0 elsewhere). For example, $\Delta^1$ has vertices $x_0=(1,0)$ and $x_1=(0,1)$, and $\Delta^2$ has vertices $x_0=(1,0,0)$, $x_1=(0,1,0)$, $x_2=(0,0,1)$. We require that the simplex map $\Delta(F)$ send the vertices to the chosen points on the corresponding strata, i.e.
\begin{equation}
    \begin{aligned}
    \label{s}
        \Delta(F)(x_i)=x(L^i),\quad  \Delta(F)(0)=x(U).
    \end{aligned}
\end{equation}
where $U$ denotes the ambient open cell and the barycenter choice is used to fix the interior image.

For every flag $F(L^1 \supset L^2\supset \cdots \supset L^p)$ define a map
\begin{equation}
    \begin{aligned}\label{c(F)}
        c(F): (S^1)^p \rightarrow U(\xi).
    \end{aligned}
\end{equation}
which gives the local product of small circles around the hyperplanes in the flag. The duality relation between ordered hyperplanes and flags is encoded by the integral pairing
\begin{equation}
    \begin{aligned}
    \label{D}
       \langle (H_1,\cdots,H_p),F\rangle=(-1)^{\vert \sigma \vert } \int_{c(F)} \iota(H_1,\cdots,H_p)=(-1)^{\vert \sigma \vert}.
    \end{aligned}
\end{equation}
where $ (H_1,\dots,H_p)$ is ordered according to some permutation $\sigma$ of the hyperplanes.

We introduce the local system (twisted)
\begin{equation}
    \ell^{\alpha^{\prime} a}=\prod_{H} \exp^{\left(\alpha^{\prime} a(H)\,log\,f_H\right)} ,
\end{equation}
which associates to each hyperplane $H$ a weight $a(H)$ through its defining function $f_H$. According to Lemma~(4.73) of Schechtman et al\cite{Schechtman}, the integral of logarithmic forms over a flag simplex with respect to this local system admits the asymptotic expansion
\begin{equation}
    \begin{aligned}
    \label{l}
    \int_{\vartriangle (F)} \iota(H_1\cdots H_p )\ell^{\alpha^{\prime} a(F)}=\frac{A}{{\alpha^{\prime}}^p}+O(\alpha^{\prime^{-p+1}})\quad (\alpha^{\prime}\rightarrow 0).
    \end{aligned}
\end{equation}
for some $\sigma\in S_p$ with flag $F=F(H_{\sigma (1)}, \cdots, H_{\sigma (p)})$, one has
\begin{equation}
    \begin{aligned}
        A=\frac{(-1)^{|\sigma|}}{a(F)},\qquad a(F)=\prod_{i=1}^p a(H_{\sigma(i)}),
    \end{aligned}
\end{equation}
and $A=0$ otherwise. This statement is directly related to the leading-order behavior of $Z$‑theory amplitudes in the $\alpha^{\prime} \to 0$ limit, as obtained from generalized Pochhammer contours \cite{Mizera_2017}.
In this limit, the result is equivalent to the regularization of a flag simplex: 
 \begin{equation}
    \begin{aligned}\label{reg}
       \text{reg} \, \Delta F=\frac{c(F_m)}{(\alpha^{\prime})^p a(F_m)}.
    \end{aligned}
\end{equation} 
where $p$ denotes the degree of the flag, and $c(F_m)$ represents the corresponding integration cycle associated with the flag $F_m$ (see Eq.~(\ref{c(F)})). An analogous form of regularization also applies to twisted cycles in the $\alpha^{\prime} \to 0$ limit \cite{Yoshida}.

\textbf{Four‑point example}. Applying the lemma to the four-point case yields the leading asymptotic behaviour
\begin{equation}
    \begin{aligned}
        &\int_{\Delta (F_1)} \iota(H_{12})\ell^{\alpha^\prime a}=\frac{1}{\alpha^\prime s}+O({\alpha^\prime}^0),\\ 
        &\int_{\Delta (F_2)} \iota(H_{23})\ell^{\alpha^\prime a}=\frac{1}{\alpha^\prime t}+O({\alpha^\prime}^0).
    \end{aligned}
\end{equation}
Under the regularization described above, one finds schematically
\begin{equation}
    \text{reg} \,\Delta F_1 =\text{reg}\,(0,1/2)=\frac{c(0,\varepsilon)}{ s\alpha^\prime}, \quad \text{reg} \, \Delta F_2 =\text{reg}\,(1/2,1)=\frac{c(1-\varepsilon ,1)}{ t\alpha^\prime}.
\end{equation}
where $c(0,\varepsilon)$ and $c(1-\varepsilon ,1)$ denote the small loops around the singular points at $0$ and $1$, with opposite orientations, respectively \cite{Yoshida}.

Combining the flag regularization~\eqref{reg} with the duality relation~\eqref{D}, we obtain for the two degree-one flags
\begin{equation}
    \begin{aligned}
        \lim\limits_{\alpha^\prime \to 0} \int_{\bigtriangleup(F_1)}\iota(H_{12}) \ell^{\alpha^\prime a}= \frac{1}{\alpha^\prime s } \int_{c(F_1)}\iota(H_{12})=\frac{1}{\alpha^\prime s}.\\ 
        \lim\limits_{\alpha^\prime \to 0} \int_{\bigtriangleup(F_2)}\iota(H_{23}) \ell^{\alpha^\prime a}= \frac{1}{\alpha^\prime t } \int_{c(F_2)}\iota(H_{23})=\frac{1}{\alpha^\prime t}.
    \end{aligned}
\end{equation}

The sum of the this result is equivalent to the integral representation of the twisted cycle and cocycle in the $\alpha^{\prime}\to0$ limit\cite{Mizera_2017}:
\begin{equation}
    \lim_{\alpha^{\prime}\to0}\langle C(1234),PT(1234)\rangle =\frac{1}{\alpha^{\prime} s}+\frac{1}{\alpha^{\prime} t},
\end{equation}
where the Parke–Taylor form is $PT(1234) = \iota(H_{12}) - \iota(H_{23})$, and the integral over $[0,1]$ (cycle $C(1234)$) is parameterized by the flag simplices as
\begin{equation}
    \int_{0}^{1} = \int_{\Delta (F_1)-\Delta(F_2)}.
\end{equation}
The full integral thus reduces to the sum of two local contributions. The negative sign in $\iota(H_{23})$ is exactly canceled by the opposite orientation of the flag simplex $\Delta(F_2)$, ensuring that both local terms contribute positively to the final result.

For a different integral, such as
\begin{equation}
    \lim_{\alpha^{\prime}\to0}\langle C(1234),PT(1324)\rangle=-\frac{1}{\alpha^{\prime}t}
\end{equation}
the result arises because $\iota (H_{23})$ in $PT(1324)$is dual only to the flag $F_2(H_{23})$ in $C(1234)$, while the contributions from the other flags vanish.

\textbf{Five-point example}. In a convenient affine patch (for instance with $z_1=0,\ z_4=1,\ z_5=\infty$) the relevant hyperplanes are 
\begin{equation}
    \begin{aligned}
H_{12}:z_2=0 ,\,
H_{24}:z_2-1=0,\,
H_{34}:z_3-1=0,\,
H_{13}:z_3=0,\,
H_{23}:z_2-z_3=0.
    \end{aligned}
\end{equation}
\begin{figure}[htbp]\label{fig:blow-up}
\centering
\includegraphics[width=.4\textwidth]{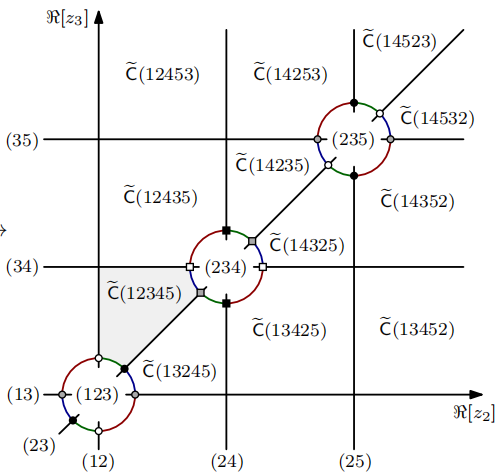}
\qquad
\includegraphics[width=.4\textwidth]{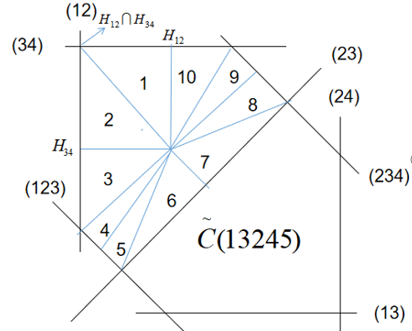}
\caption{The left figure, taken from \cite{Mizera_2017}, shows all twisted cycles $\hat{C}(\alpha)$ of the Riemann space $\mathcal{M}_{0,5}$ after the blow‑up.  
The right figure decomposes the integral region $\hat{C}(12345)$ into 10 simplices according to equation~(\ref{s}).}
\end{figure}
After blowing up \cite{Mimachi_2003} the neighbourhoods of the corners $(0,0)$ and $(1,1)$ and performing a barycentric subdivision, the integration region $\widetilde C(12345)$ decomposes into ten 2‑simplices.
For example, the flag simplex $\Delta F_1(H_{12}\supset H_{12}\cap H_{34})$maps the point $x_1$ onto a codimension‑1 edge (the hyperplane $H_{12}$) and the point $x_2$ onto a codimension‑2 edge (the intersection point of the two hyperplanes $H_{12}$ and $H_{34}$). Similarly, $\Delta F_2(H_{34}\supset H_{34}\cap H_{12})$ maps the point $x_2$ onto a codimension‑1 edge (the hyperplane $H_{34}$) and the point $x_2$ onto a codimension‑2 edge (the intersection $H_{34} \cap H_{12}$). Applying the lemma to the flag simplices and summing the leading contributions yields (see Appendix \ref{appendix})
\begin{equation}
    \begin{aligned}
    \label{A}
\frac{1}{\alpha^{\prime^2 }}\left(\frac{1}{s_{12}s_{123}}+\frac{1}{s_{23}s_{123}}+\frac{1}{s_{34}s_{234}}+\frac{1}{s_{23}s_{234}}+\frac{1}{s_{12}s_{34}}\right).
    \end{aligned}
\end{equation}
We replace the weight of each flag by the product of the kinematic invariants $s_{ij}s_{ijk}$. This expression agrees with the field‑theory limit of the Z‑amplitude $\langle C(12345),PT(12345)\rangle$.

Consider the Parke–Taylor form 
\begin{equation}
    \begin{aligned}
PT(13245)=\iota (H_{32},H_{24})- \iota (H_{13},H_{24})+\iota (H_{13},H_{32}).
    \end{aligned}
\end{equation}
The first term, $\iota(H_{32},H_{24})$, and the last term, $\iota(H_{13},H_{32})$, can each be decomposed into four contributions, as in Eq.~(\ref{32}). Among these, the only non-zero contributions come from the overlapping regions of the integral cycles $C(12345)$ and $\widetilde C(13245)$, corresponding to two flag simplices whose vertices are located at the points $(0,0)$ and $(1,1)$ in the chosen coordinate system, respectively.

For $PT(12345)$, according to the decomposition of flag simplices, only simplices 6 and 7 contribute. Therefore, using Eq.~(\ref{l}) and Eq.~(\ref{32}), we find that at leading order in $\alpha^{\prime}\to 0$ limit,
\begin{equation}
    \begin{aligned}
        \lim_{\alpha^{\prime} \rightarrow 0 }\langle C(12345),PT(13245)\rangle &=\frac{1}{\alpha^\prime a(H_{23})} \int_{c_{\vert H_{23}}}(\frac{1}{\alpha^\prime a(L^2_1)}\int_{c(L^2_1)} \iota(H_{13},H_{23})+\frac{1}{\alpha^\prime a(L^2_2)}\int_{c(L^2_2)} \iota(H_{23},H_{24}))\notag\\ 
&=\frac{1}{(\alpha^\prime)^2}\left(\frac{1}{a(H_{23}) a(L^2_1)} +\frac{1}{a(H_{23}) a(L^2_2)}\right)
=\frac{1}{(\alpha^\prime)^2} \left(\frac{1}{s_{23} s_{123}} +\frac{1}{s_{23} s_{234}}\right).
    \end{aligned}
\end{equation}
which is in agreement with the result obtained by Mizera\cite{Mizera_2017}. 

Here, the quantities $a(H_{23}) a(L^2_1)$ and $a(H_{23}) a(L^2_2)$ denote the corresponding weights of the flags $F(H_{23} \supset L_1^2)$ and $F(H_{23} \supset L_2^2)$, with $L_1^2$ and $L_2^2$ representing the points obtained from the intersections of $H_{23}$ with $H_{123}$ and $H_{234}$, respectively. For this flag simplex parameterization, the integral can be expressed as
\begin{equation}
\begin{aligned}
    \int_{c(F)} 
    = \int_{c(F_1)} + \int_{c(F_2)}= \int_{\,c|_{H_{23}}} \Big( \int_{c(L^2_1)} + \int_{c{(L^2_2)}} \Big).
\end{aligned}
\end{equation}
Similarly, other choices of twisted cycles and cocycles yield consistent results for their corresponding integrals.

\textbf{For the general case}, the lemma can be interpreted as a pairing $ \mathcal{A}^p \times \mathcal{F}^p  \rightarrow \mathbb{C} $ between the space of $p$-forms $\mathcal{A}^p$ and the space of $p$-dimensional flags $\mathcal{F}^p$. This pairing yields the same result as the Z‑amplitude $Z_{\beta}(\gamma)$ in the field‑theory limit, corresponding to the pairing of twisted cycles and twisted cocycles \cite{Mizera_2017}.
In our method, we can view this relationship as represented by flag
\begin{equation}
    \begin{aligned}
        \lim_{\alpha^{\prime} \rightarrow 0 }\langle C(\alpha),PT(\beta)\rangle=\frac{1}{(\alpha^\prime)^p}\sum_m\frac{\langle F_m,PT(\beta)\rangle}{ S^{(a)}(F_m,F_m)},
    \end{aligned}
\end{equation}
where the sum runs over all flags $F_m$ generated by the arrangement of hyperplanes that define the integration domain $C(\alpha)$.

In summary, the regularization via flag simplices isolates the dominant contributions in the limit $\alpha^{\prime} \rightarrow 0$, and reproduces the same leading behaviour as obtained from the generalized Pochhammer contour construction\cite{Mizera_2017}. Both procedures therefore yield, at leading order, the well‑known bi‑adjoint scalar amplitudes.

\subsection{CHY Formula and Special Flag }
The master function associated to the weights $a(H)$ is the multivalued function on $U(\xi)$ defined by
\begin{equation}\label{master}
    \phi \;=\; \sum_i a_i \,\log f_{H_i},
\end{equation}
where $f_{H_i}$ denotes a defining polynomial for the hyperplane $H_i$. Taking 
\begin{equation}
    f_{H_{ij}} = z_i - z_j,
\end{equation}
and fixing the $\mathrm{SL}(2,\mathbb{C})$ gauge, the critical points of $\phi$ coincide with the solutions of the scattering equations. The number of critical points equals the Euler characteristic of $U$, denoted $\lvert\mathcal{X}(U)\rvert$; for the moduli space of genus‑zero Riemann spheres with $n$ punctures one has $\lvert\mathcal{X}(U)\rvert=(n-3)!$, see, e.g.,\cite{10.2969/jmsj/03920191,Mizera_2020A,Varchenko_2011}.

The logarithmic 1-form associated with the master function $\phi$ is
\begin{equation}
    \omega \;=\; \sum_j a(H_j)\frac{df_{H_j}}{f_{H_j}},
\end{equation}
and critical points of $\phi$ are determined by the condition $\omega|_{u}=0$.

For a regular function $f$ on $U(\xi)$, the Grothendieck residue at an isolated critical point $u$ is given by
\begin{equation}\label{grothendieck}
    \operatorname{Res}_u (f) \;=\; \frac{1}{(2\pi i)^p}\int_{\Gamma_u} \frac{f(z)\,dz_1\wedge\cdots\wedge dz_p}
    {\prod_{i=1}^p \frac{\partial\phi}{\partial z_i}},
\end{equation}
where the cycle $\Gamma_u=\{z\in U_u:\ |\partial\phi/\partial z_i|=\varepsilon_i\}$ is taken with the orientation determined by
\begin{equation}
    d\arg\frac{\partial\phi}{\partial z_1}\wedge\cdots\wedge d\arg\frac{\partial\phi}{\partial z_p}\ge 0.
\end{equation}

The residue bilinear identity localised at the critical points reads
\begin{equation}\label{res-bilin}
    \frac{1}{(2\pi i)^p}\int_{\Gamma_u}\frac{f\,g\,dz_1\wedge\cdots\wedge dz_p}
    {\prod_{i=1}^p \frac{\partial\phi}{\partial z_i}}
    \;=\; \sum_{u}\frac{f(u)\,g(u)}{\operatorname{Hess}(u)},
\end{equation}
with the Hessian
\begin{equation}
    \operatorname{Hess}(u)=\det\!\big(\partial_{z_i}\partial_{z_j}\phi(u)\big).
\end{equation}

Consider a top-degree differential form
\begin{equation}
    H=f_Hdz_1\wedge\cdots\wedge dz_p\in\mathcal{A}^p.
\end{equation}
The specialization map
\begin{equation}
    F: U \longrightarrow \mathcal{F}^p,\qquad u\mapsto F(u),
\end{equation}
is defined by
\begin{equation}
    \langle H,F(u)\rangle=f_H(u).
\end{equation}
The vectors F(u) span the subspace $Sing\mathcal{F}^p\subset\mathcal{F}^p$ (the space annihilating the image of $\nabla_\omega:A^{p-1}\to A^p$). Writing a basis of flags ${F_m}$ with dual top‑forms $H^m=f^mdz_1\wedge\cdots\wedge dz_p$, one has the expansion \cite{Varchenko_2011}
\begin{equation}\label{special-decomp}
    F(u) \;=\; \sum_m f^m(u)\,F_m.
\end{equation}

The contravariant form $S^{(a)}$ satisfies relation\cite{varchenko2016critical} (up to the sign convention)
\begin{equation}\label{Hess-rel}
    S^{(a)}\big(F(u),F(u)\big) \;=\; (-1)^p\,\operatorname{Hess}(u).
\end{equation}
Different critical points give mutually orthogonal special vectors with respect to $S^{(a)}$:
\begin{equation}
    S^{(a)}\big(F(u_1),F(u_2)\big)=0\quad (u_1\ne u_2).
\end{equation}

Define the linear map $\varepsilon$ on regular functions by
\begin{equation}
    \varepsilon(g) \;=\; \sum_u \sum_m \frac{g(u)\,f^m(u)}{\operatorname{Hess}(u)}\,F_m \;=\; \sum_u \frac{g(u)F(u)}{\operatorname{Hess}(u)}.
\end{equation}
This map identifies the residue pairing and the contravariant form on $\operatorname{Sing}\mathcal{F}^p$ up to the sign $(-1)^p$:
\begin{equation}
    (f,g) \;=\; (-1)^p\, S^{(a)}\big(\varepsilon(f),\varepsilon(g)\big).
\end{equation}

Hence for two top forms (e.g. Parke–Taylor forms) $PT(\alpha)=f_\alpha\,dz_1\wedge\cdots\wedge dz_p$ and $PT(\beta)=f_\beta\,dz_1\wedge\cdots\wedge dz_p$, with the help of \eqref{Hess-rel}one obtains
\begin{equation}\label{CHYlocal}
    \langle PT(\alpha),PT(\beta)\rangle
    \;=\;
    \sum_u \frac{\langle PT(\alpha),F(u)\rangle\,\langle F(u),PT(\beta)\rangle}{S^{(a)}\big(F(u),F(u)\big)}
    \;=\;
    \sum_u \frac{f_\alpha(u)\,f_\beta(u)}{S^{(a)}\big(F(u),F(u)\big)},
\end{equation}
Here, $\langle PT(\alpha),F(u)\rangle = f_{\alpha} (u)$ and $\langle PT(\beta),F(u)\rangle =f_{\beta}(u)$. This expression coincides precisely with the CHY localisation formula for the bi-adjoint scalar amplitude on the solutions of the scattering equations~\cite{Cachazo_2014563,Fu_2020}.

\section{Conclusion}\label{section 5}

In this work, we first identify each standard flag with a trivalent Feynman diagram on the moduli space $\mathcal{M}_{0,n}$, so that the propagators of the diagram are realized as bilinear forms of the flag under the contravariant map $S^{(a)} (F_m,F_m)$. The geometric identification of flags with Feynman diagrams ensures that the residue-theorem realization of the Bern–Carrasco–Johansson (BCJ) duality is automatically satisfied through the gap relation between a flag and its gapped counterpart. Using the duality between the Orlik–Solomon algebra and flags, we express intersection numbers of twisted cocycles in terms of flags, which naturally reproduce the bi-adjoint scalar amplitude. By introducing flag simplices, the Z-theory amplitude emerges in the $\alpha^{\prime}\rightarrow 0$ limit as a sum over all flag simplices, reflecting its algebraic structure in terms of pairings between twisted cycles and cocycles and its relation to associahedra via Catalan-number vertices\cite{Mizera_2017}. Mapping the critical points of the master equation to special flags $F(u)$ yields a subspace whose dimension matches the Euler characteristic, $\vert \mathcal{X}(U)\vert= (n-3)!$, and coincides with the solutions of the scattering equations when hyperplanes $z_i-z_j=0$ are weighted by Mandelstam invariants $s_{ij}$, giving rise to the CHY representation of the amplitude. Together, these results establish a unified framework connecting flag geometry, intersection theory, and scattering amplitudes in both field-theory and string-theory limits.

For the massless case, we choose the weight of each hyperplane to be $s_{ij}$  the above results also hold under the shift
$s_{ij} \rightarrow s_{ij} + \varLambda $ for all equal external masses $\varLambda $ \cite{Mizera_2020}. Finally, it is worth exploring a generalized Orlik–Solomon algebra and flag framework that incorporates color factors in Yang–Mills theory; in such a case, our formulas can be extended to yield results for more general gauge theories.
\begin{acknowledgments}
I am grateful to my supervisor Chih-Hao Fu for his help during my Postgraduate studying. I also thank to Yihong Wang for valuable discussions and Chang Hu for useful suggestion.
\end{acknowledgments}

\appendix

\section{The Detail of Equation (\ref{A})}\label{appendix}
In this appendix, we present in detail the five‑point example of the lemma from Section~\ref{subsection 4}, using the barycentric subdivision of a flag simplex. 
The result agrees with the $Z$‑theory amplitude in the $\alpha^{\prime} \to 0$ limit obtained via the generalized Pochhammer contour, and is directly related to the concept of the associahedron arising from the blow‑up of the moduli space. For flag simplex 1 in Figure~\eqref{fig:blow-up}, we have
\begin{equation}
    \begin{aligned}
        F_1=F(H_{12}\supset H_{12}\cap H_{34}),
    \end{aligned}
\end{equation}
with corresponding weight
\begin{equation}
    \begin{aligned}
        a(F_1)=a(H_{12})(a(H_{12})+a(H_{34})),
    \end{aligned}
\end{equation}
Similarly, for flag simplex~2,
\begin{equation}
    \begin{aligned}
        F_2=F(H_{34}\supset H_{34}\cap H_{12}),
    \end{aligned}
\end{equation}
with corresponding weight
\begin{equation}
    \begin{aligned}
        a(F_2)=a(H_{34})(a(H_{34})+a(H_{12})).
    \end{aligned}
\end{equation}
From the lemma we have
\begin{equation}
    \begin{aligned}
       & \int_{\bigtriangleup F_1} \iota(H_{12},H_{34}) \ell^{\alpha^{\prime } a} = \frac{1}{\alpha^{\prime 2 } a(F_1)},\\ 
       & \int_{\bigtriangleup F_2} \iota(H_{12},H_{34}) \ell^{\alpha^{\prime } a} = \frac{-1}{\alpha^{\prime 2 } a(F_2)},
    \end{aligned}
\end{equation}
where the minus sign arises because $F_2=F(H_{34},H_{12})$ corresponds to an odd permutation of $F_1=F(H_{12},H_{34})$.  
Geometrically, the two simplices are oriented in opposite directions—one clockwise and the other counterclockwise—under the map from flag simplices to the integral region (orientation defined as $x(U) \to x(L^1) \to x(L^2)$).  
Thus the total contribution is
\begin{equation}
    \begin{aligned}
        \int_{\bigtriangleup F_1- \bigtriangleup F_2 } \iota(H_{12},H_{34}) \ell^{\alpha^{\prime } a} = \frac{1}{\alpha^{\prime 2}} (\frac{1}{a(F_1)}+\frac{1}{a(F_2)}) = \frac{1}{\alpha^{\prime 2}} \frac{1}{a(H_{12})a(H_{34})}.
    \end{aligned}
\end{equation}
For each pair of consecutive flag simplices $(\Delta F_i, \Delta F_{i+1})$, the mapped integral regions have opposite orientations, cancelling the $-1$ factor from the permutation in the lemma.  

Applying the same reasoning:
- For flag simplices 3 and 4:
\begin{equation}
    \begin{aligned}
        F_3=F(H_{12}\supset H_{123}\cap H_{12}), \quad  & a(F_3)=a(H_{12})(a(H_{12})+a(H_{123})),\\
        F_4=F(H_{123}\supset H_{123}\cap H_{12}), \quad & a(F_4)=a(H_{123})(a(H_{123})+a(H_{12})).
    \end{aligned}
\end{equation}
By the lemma, the contribution from $\iota(H_{12},H_{123})$ is
\begin{equation}
    \begin{aligned}
        \frac{1}{\alpha^{\prime 2 }} \frac{1}{a(H_{12})a(H_{123})}.
    \end{aligned}
\end{equation}
- For flag simplices 5 and 6:
\begin{equation}
    \begin{aligned}
        F_5=F(H_{123}\supset H_{123}\cap H_{23}),\quad & a(F_5)=a(H_{123})(a(H_{123})+a(H_{23})), \\
        F_6=F(H_{23}\supset H_{23}\cap H_{123}),\quad & a(F_6)=a(H_{23})(a(H_{23})+a(H_{123})).
    \end{aligned}
\end{equation}
By the lemma, the contribution from $\iota(H_{23},H_{123})$ is 
\begin{equation}
    \begin{aligned}
         \frac{1}{\alpha^{\prime 2 }} \frac{1}{a(H_{23})a(H_{123})}.
    \end{aligned}
\end{equation}
- For flag simplices 7 and 8:
\begin{equation}
    \begin{aligned}
        F_7=F(H_{23}\supset H_{23}\cap H_{234}),\quad & a(F_7)=a(H_{23})(a(H_{23})+a(H_{234})),\\ 
        F_8=F(H_{234}\supset H_{234}\cap H_{23}),\quad &  a(F_8)=a(H_{234})(a(H_{23})+a(H_{234})).
    \end{aligned}
\end{equation}
By the lemma, the contribution from $\iota(H_{23},H_{234})$ is
\begin{equation}
    \begin{aligned}
         \frac{1}{\alpha^{\prime 2}} \frac{1}{a(H_{23})a(H_{234})}.
    \end{aligned}
\end{equation}
- For flag simplices 9 and 10:
\begin{equation}
    \begin{aligned}
        F_9=F(H_{234}\supset H_{234}\cap H_{34}),\quad & a(F_9)=a(H_{234})(a(H_{234})+a(H_{34})),\\ 
        F_{10}=F(H_{23}\supset H_{23}\cap H_{123}),\quad & a(F_{10})=a(H_{34})(a(H_{34})+a(H_{234})).
    \end{aligned}
\end{equation}
By the lemma, the contribution from $\iota(H_{34},H_{234})$ is
\begin{equation}
    \begin{aligned}
         \frac{1}{\alpha^{\prime 2}} \frac{1}{a(H_{34})a(H_{234})}.
    \end{aligned}
\end{equation}

Finally, noting that $PT(12345)$ decomposes into five terms, each dual to a corresponding flag, we sum over all ten contributions and replace the weights $a(H_i)$ by the Mandelstam variables, with 
$$s_{123} = s_{12} + s_{23} + s_{13}, \quad s_{234} = s_{23} + s_{34} + s_{24}.$$  Summing all the above contributions, we obtain 
the $Z$‑theory amplitude in the $\alpha^{\prime} \to 0$ limit:
\begin{equation}
    \begin{aligned}
     \lim_{\alpha^{\prime}\to 0 } \langle C(12345),PT(12345) \rangle = \frac{1}{\alpha^{\prime 2 }} \left(\frac{1}{s_{12}s_{123}} +\frac{1}{s_{23}s_{123}}  +\frac{1}{s_{34}s_{234}}  +\frac{1}{s_{23}s_{234}}  +\frac{1}{s_{12}s_{34}}\right) .  
    \end{aligned}
\end{equation}

\nocite{*}
\bibliography{flag}

\end{document}